\documentclass[twocolumn,nofootinbib]{revtex4}
\usepackage{amsmath,amssymb}
\usepackage{bbm}

\newcommand{\comment}[1]{}
\newcommand{\eqa}{\begin{eqnarray}}
\newcommand{\neqa}{\end{eqnarray}}
\newcommand{\be}{\begin{equation}}
\newcommand{\ee}{\end{equation}}

\renewcommand{\texttt}{{}}
\usepackage{graphicx}

\begin{document}
\title{\large Fractal Space\hspace{0.05cm}-\hspace{-0.05cm}Time from Spin-Foams}
\author{Elena Magliaro$^{a}$, Leonardo Modesto$^{b}$ and Claudio Perini$^{ac}$} %
\affiliation{${}^a$Centre de Physique Th\'eorique de Luminy, Case 907, F-13288 Marseille, EU}
\affiliation{${}^b$Perimeter Institute for Theoretical Physics, 31 Caroline St., Waterloo, ON N2L 2Y5, Canada}
\affiliation{${}^c$Dipartimento di Matematica, Universit\`a degli Studi Roma Tre, I-00146 Roma, EU}


\date{\small\today}

\begin{abstract} \noindent
In this paper we perform the calculation of the spectral dimension of spacetime in 4d quantum gravity
using the Barrett-Crane (BC) spinfoam model. 
We realize this 
considering a very simple decomposition of the 4d spacetime already used in the graviton 
propagator calculation and we  
introduce a boundary state which selects a classical geometry on the boundary.
We obtain that
the spectral dimension of the spacetime runs from $\approx 2$ to $4$ 
when the energy of a probe scalar field decreases from high $E \lesssim E_P/25$ to low energy. 
The spectral dimension at the Planck scale $E \approx E_P$ depends on the areas spectrum
used in the calculation. For three different spectra $l_P^2 \sqrt{j(j+1)}$, $l_P^2 (2 j+1)$ 
and $l_P^2 j$ we find respectively dimension $\approx 2.31$, $2.45$ and $2.08$.

\end{abstract}
\maketitle
%
\paragraph*{Introduction.} 
In past years many approaches to quantum gravity studied the fractal properties of quantum spacetime. 
In particular in {\em causal dynamical triangulation} (CDT) \cite{cdt} 
and {\em asymptotically safe quantum gravity} (ASQG) \cite{asqg}, a fractal analysis 
of spacetime gives a two dimensional effective manifold
at high energy. In both approaches the spectral
dimension is ${\mathcal D}_s=2$ at small scales and ${\mathcal D}_s=4$ at large scales. The previous ideas have been applied also in the context of {\em noncommutative geometry} to a quantum sphere and
$\kappa$-Minkowski \cite{dario}, in {\em causal sets} \cite{DR1}\cite{DR2} and in loop quantum gravity \cite{Modesto:2008jz}. See \cite{Carlip:2009kf} for a summary and new insights.
Spectral properties have been considered also for the cosmology of a Lifshitz universe
\cite{calcagni}. Spectral analysis is a useful tool to understand the {\em effective form}
 of space at small and large scales. 
We believe that the fractal analysis could be also a useful tool to predict the behaviour 
of $n$-point correlation functions at small scales \cite{propagatoreR}\cite{propagatoreMR} and to attack the 
singularity problems of general relativity in the full theory of quantum gravity \cite{CBH-AB}\cite{CBH-M}. 

In this paper we apply  to the Barrett-Crane (BC) spin foam model \cite{barrett} for Riemannian 4d quantum gravity the analysis introduced in \cite{Modesto:2008jz}. The fractal properties of 3d spacetime were studied in \cite{fractal3d}. The main ingredient is the general boundary formalism 
which provides quantum amplitudes associated to a finite spacetime region for a given boundary 3-geometry \cite{gb1}\cite{gb2}. The formalism is implemented in the same spirit of the calculation of LQG graviton propagator \cite{propagatoreMR}\cite{propagatoreR}\cite{propagatoreBMRS}\cite{propagatoreBMP}\cite{propagatore3d}\cite{propagatoreAR1}\cite{propagatoreAR2}\cite{propagatoreABR3}.
For our purposes we will consider a  boundary state peaked over the boundary geometry of a single $4$-simplex.

The paper is organized as follows. In the first section we define the framework and recall the definition of spectral dimension 
in diffusion processes.
In the second section we define the spectral dimension  
in quantum gravity.
The analysis is general and not strongly related to the 
specific models. 
We continue 
calculating explicitly the spectral dimension for the BC theory using
the general boundary formalism to define the 4d quantum gravity path integral. 

\paragraph{Spectral dimension in diffusion processes.} 
The following definition of fractal dimension is borrowed
from the theory of diffusion processes on fractals \cite{avra} and easily 
adapted to the
quantum gravity context. 
Let us study the Brownian motion of a test particle moving on a $d$-dimensional Riemannian manifold $\mathcal M$ with a fixed smooth metric $g_{\mu\nu}(x)$.
The probability density for the particle to diffuse from $x'$ to $x$ during the fictitious time 
(this is just a fictitious time since we are probing the spacetime properties, not only the properties of space) 
 is the heat-kernel $K_g(x,x';T)$, which satisfies the heat equation
\begin{eqnarray}
\label{heateq}
\partial_T K_g(x,x';T)=\Delta_g K_g(x,x';T)
\end{eqnarray}
where $\Delta_g$ denotes the covariant Laplacian: 
\begin{eqnarray}
\Delta_g\phi\equiv \frac{1}{\sqrt{g}} \,\partial_\mu(\sqrt{g}\,g^{\mu\nu}\,\partial_\nu
\phi). 
\end{eqnarray}
The heat-kernel is a matrix element of the operator 
$\exp(T\,\Delta_g)$, acting on the real Hilbert space $L^2(\mathcal M, \sqrt{g} \, \text{d}^d x)$, between position eigenstates
\begin{eqnarray}
K_g(x,x';T) =\langle x^{\prime} | \exp(T\,\Delta_g) | x \rangle.
\label{EK}
\end{eqnarray} 
Its trace per unit volume,
\begin{eqnarray}
\label{trace}
&& P_g(T)\equiv V^{-1}\int \text{d}^dx\,\sqrt{g(x)}\,K_g(x,x;T) \nonumber \\
&& \hspace{1cm} \equiv 
V^{-1}\,{\rm Tr}\,
\exp(T\,\Delta_g)\;,
\end{eqnarray}
has the interpretation of an average return probability. Here $V\equiv\int
d^dx\,\sqrt{g}$ denotes the total volume. It is well known that $P_g(T)$
possesses an asymptotic expansion for $T\rightarrow 0$ of the form
$P_g(T)=(4\pi T)^{-d/2}\sum_{n=0}^\infty A_n\,T^n$. The coefficients $A_n$ have a geometric meaning, i.e. $A_0$ is the volume of the manifold and if $d=2$ then $A_1$ is proportional to the Euler characteristic. From the knowledge of the function $P_g$ one can recover the dimensionality of the
manifold as the limit for small $T$ of
\begin{eqnarray}
\label{dimform}
\mathcal D_s\equiv-2\frac{\text{d}\ln P_g(T)}{\text{d}\ln T}.
\end{eqnarray}
If we consider arbitrary fictitious times $T$, this quantity may depend on the scale we are probing. \eqref{dimform} is the definition of fractal dimension we will use.
 \paragraph{Spectral dimension in quantum gravity.}
In quantum gravity it is natural to
replace the Laplacian in the heat-kernel equation (\ref{heateq}) with an effective Laplacian associated to the expectation value of the operator $\langle \Delta_g \rangle$. It is defined by the path integral
\begin{eqnarray}
\label{pexpect}
\langle \Delta_{\hat{g}} \rangle\equiv 
\int_{\Psi} Dg \, \Delta_{g} \, e^{iS(g)}.
\end{eqnarray}
 
Using the effective Laplacian defined above the heat-kernel equation (\ref{heateq}) 
becomes 
\begin{eqnarray}
\label{heateqs}
 \partial_T K(x,x';T)  = 
\langle  \Delta_{\hat{g}} \rangle K(x,x';T) . 
\end{eqnarray}
Given $\langle \Delta_{\hat{g}} \rangle$ we can define the average return probability 
$P(T)$ in analogy with the classical definition, 
\begin{eqnarray}
P(T) = V^{-1}\,{\rm Tr}\,
\exp(T\, \langle \Delta_{\hat{g}} \rangle)\;.
\label{EKPT}
\end{eqnarray} 
The spectral dimension of quantum 
spacetime is defined as in (\ref{dimform}). 
\comment{
In particular we will find for the effective Laplacian the form
$\langle \Delta_{\hat{g}} \rangle = (-1)^{z+1} \Delta^z$, 
where $z=z(k)$ is a function of the energy scale $k$. The coefficient $(-1)^{z+1}$
is necessary to have an elliptic operator. For a generic value of the function $z(k)$
equation (\ref{heateqs}) can be rewritten in the form 
\begin{eqnarray}
\label{heateqsz}
 \partial_T K(x,x';T)  =  (-1)^{z+1} \Delta^z  K(x,x';T). 
\end{eqnarray}
The solution of  (\ref{heateqsz}) in Fourier transform and general dimension $d$ is locally given by 
\begin{eqnarray}
\label{solheateqsz}
K(x,x';T)  =  \int \frac{\text d^d k }{(2 \pi)^d} e^{i k_i (x_i - x'_i)} \, e^{ - T |k|^{2 z}}.
\end{eqnarray}
The trace of the heat kernel for $x^{\prime} \approx x$ is
\begin{eqnarray}
\label{trK}
P(T)  =  \frac{\rm const.}{T^{\frac{d}{2 z}}}
\end{eqnarray}
and the spectral dimension can be computed by (\ref{specdim}) in terms of the scaling 
function $z(k)$: ${\mathcal D}_s = d/z$.
}
\paragraph{Including quantum gravity effects.} We expect the effective Laplacian to run with the probed energy scale $k$ when quantum gravity effects become important. Above some energy scale the quantum spacetime reveals its fuziness. So we expect that
\begin{equation}
\langle\Delta_{\hat g}\rangle=\Delta_k=F(k)\Delta_{k_0}
\end{equation}
where $\Delta_{k_0}$ is the Laplacian at a reference infrared momentum $k_0$, and $F(k)$ is a scaling function of the energy scale. For instance, in the case the i.r. Laplacian is the flat one, we have, expading in Fourier modes:
\begin{equation}
P(T)=\int\text d^d p\,e^{T F(k)p^2}.
\end{equation}
We argue that the scales $k$ and $p$ should be physically identified, since $p$ represents the energy probed by the $p-$th mode of a scalar field. Then we define an average return probability which includes quantum gravity effects by
\begin{equation}\label{TPKwithQG}
P_\text{QG}(T)\equiv\int\text d^d p\,e^{T F(p) p^2}.
\end{equation}
If, as we will find,
$F(k)\propto k^{2(z-1)}$, 
then
\begin{equation}
P_\text{QG}(T)=\int\text d^dp\,e^{T p^{2z}}\propto\frac{1}{T^{\frac{d}{2z}}}
\end{equation}
and the fractal dimension is
$\mathcal D_s= d/z$.

\paragraph{Spectral dimension in spin foams.}
In the context of spin foam models for quantum gravity we can give a precise meaning to the formal path integral (\ref{pexpect}) and define the spectral dimension in the general boundary formalism. We introduce a gaussian state $|\psi_{\bf q} \rangle$ peaked on 
the boundary geometry
${\bf q} =(q, p)$ defined by the 3-metric and the conjugate momentum.
We can think the boundary geometry to be the boundary of a $(d-1)$-dimensional ball.
The state is Gaussian and symbolically given by:
\begin{eqnarray}
\Psi_{\bf q} (s) \sim  {\rm e}^{ -(s - q)^2 + i p  s}.
\label{gauss}
\end{eqnarray}
The amplitude (\ref{pexpect}) can be defined for a general spin foam model 
in the general boundary framework as
\begin{eqnarray}
\hspace{-0.2cm}
\frac{\langle W | \Delta_{\hat{g}} | \Psi_{\bf q} \rangle}{\langle W | \Psi_{\bf q} \rangle } = \frac{\sum_{s_1, s_2} W(s_1) \, \langle s_1 | \Delta_{\hat{g}} |s_2 \rangle \, \psi_{\bf q} (s_2)}
{\sum_s W(s) \Psi_{\bf q}(s)},
\label{pmedgb}
\end{eqnarray}
where $W(s)$ codifies the spin foam dynamics. We will consider the BC model with vertex amplitude 
$W(s)$ proportional to the $10j$-symbol $\{10 j\}$.  

The fundamental ingredient required to 
study the fractal properties of quantum spacetime is the the scaling of the metric.
To this purpose  
we consider only one simplex 
that could be part of a more complicated simplicial decomposition.
In other words we suppose spacetime to be a 3-ball with a boundary 3-sphere 
and triangulate the sphere in a very fine way (the dual of the triangulation is a spin-network).
Then we consider another 3-ball and its boundary 3-sphere but at a smaller scale. Since we are considering the 3-sphere at two different scales 
all the representations labeling the boundary spin-network will be rescaled conformally. In our simplification the 3-ball is approximated by a single 4-simplex and the boundary 
3-sphere by the ten faces of the simplex.
%
Summarizing: in 4d gravity we approximate a $3$-ball with a single simplex and
the boundary sphere $S^3$ with the surface of the simplex
given by ten triangles.

We use a conformal metric defined by $g_{\mu \nu} = \Omega^2 \, g^{o}_{\mu \nu}$, where $g^{o}_{\mu \nu}$ is a background metric  and $\Omega^2$ is the conformal factor. This is the hypothesis made in \cite{Modesto:2008jz} and \cite{fractal3d}. 
In the quantum theory the conformal factor is replaced by the area operator and the metric operator is defined by $\hat{g}_{\mu \nu} = \hat{A} \, g^{o}_{\mu \nu}$. 
To extract 
the fractal properties of spacetime we approximate the metric in the Laplacian 
with the inverse 
of the area operator in 4d. In fact the Casimir operator is related to the
area spectrum of a triangular face in the simplicial decomposition of space. 
We consider three possible area spectra
\begin{eqnarray}
A_j  = \left\{ \begin{array}{lll} 
        A_{j}^{{\rm a}} 
= l_P^2[ j(j+1)]^{\frac{1}{2}}
        \vspace{0.04cm}    \\ 
            A_{j}^{{\rm b}} = l_P^2( 2 j +1)
                  \\
                   A_{j}^{{\rm c}} = l_P^2 \, j
        \end{array} \right.
\label{length}
\end{eqnarray}
%
%
%
where $j$ is the $SU(2)$ representation dual  to the triangular face. The quantity which is necessary to compute $\langle \Delta_{\hat{g}} \rangle$ is  
\begin{equation}
\langle \hat{g}^{\mu \nu} \rangle \equiv \langle \widehat{1/A} \rangle \, g^{(0) \mu \nu}.
\end{equation}
The boundary state in the notation above is  
\begin{eqnarray}
\hspace{-0.2cm}
\Psi_{j}(j_e) \propto 
 {\rm e}^{- \frac{1}{2 j}  \sum_{e_1,e_2} M_{e_1e_2} (j_{e_1} - j)(j_{e_2} -j) + i \Phi \sum_{e} j_e}.
\label{state0}
\end{eqnarray} 
The 10 spins $j_e$ label the boundary triangles of a 4-simplex. The Gaussian is peaked over the homogeneous spin configuration, namely over an equilateral 4-simplex of size of order $\sqrt j$. The dihedral angles $\Phi = \arccos (-1/4)$ define the boundary extrinsic geometry. $M$ is a $10\times10$ matrix with positive definite real part. We chose $M=\mathbbm 1_{10\times10}$ in the numerical calculation, but other choices do not affect qualitatively the analysis.

\begin{figure}
 \includegraphics[height=4.5cm]{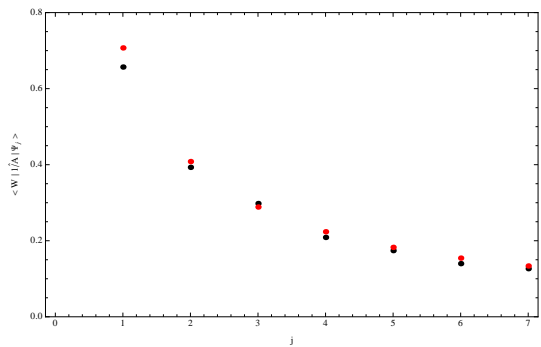}
  \caption{\label{OjT} 
 Plot of the amplitude (\ref{ampli}) against $j$ for $1 \lesssim j \lesssim 7$. The black points represent 
 the amplitude $\eta \, \langle W| \widehat{1/A^{\rm a}_{j_1}} |\Psi_j \rangle$ and the red points 
 represent $1/[j(j+1)]^{\frac{1}{2}}$.
  }
 \end{figure}

The expectation value (\ref{pmedgb}) reads
\begin{eqnarray}
&& \hspace{-0.5cm} 
\eta \, \langle W| \widehat{1/A^{\rm a}_{j_1}} |\Psi_j \rangle  =
\eta \hspace{-0.1cm}
\sum_{j_{e}=1} W(j_e) \frac{1}{A^{\rm a}_{j_1}} \Psi_{j}(j_{e})
\nonumber \\
\label{somma}
&& \hspace{-0.5cm}=\eta 
\hspace{-0.1cm}
\sum_{j_{e} =1}
\prod_{e=1}^{10} {\rm dim} ( j_e) \{10 j\} \, \frac{1}{[j_1(j_1+1)]^{\frac{1}{2}}l_P^2}
\,   \Psi_j(j_{e} ),
\label{ampli}
\end{eqnarray}
where we introduced the following notation for the normalization, 
$\eta^{-1} \equiv \langle W| \Psi_j \rangle$. 
In Fig.\ref{OjT} 
we have plotted the amplitude (\ref{ampli}) as a function of the 
$SU(2)$ representation $j$. This is also the quantum gravity scale defined by
$\ell^2 = l_P^2 j$. 

For $j\gtrsim 7$ the amplitude (\ref{ampli}) is well approximated by the classical function 
 \begin{eqnarray}
&& \hspace{-0cm} 
\eta \, \langle W| \widehat{1/A^{\rm a}_{j_1}} |\Psi_j \rangle \approx  
\frac{1}{[j(j+1)]^{\frac{1}{2}}l_P^2}.
\label{amplilargej}
\end{eqnarray}
The result (\ref{amplilargej}) can be obtained also analytically in the large $j$ limit 
replacing the amplitude $W(j)$ with the exponential of the Regge action. More precisely, since the boundary state is peaked on large values of the spins, the $10j$-symbol in the sum \eqref{somma}  can be approximated with its large spin asymptotic formula 
\begin{equation}
\label{10jRegge}
\{10 j\}\sim A(j)\cos S_{\text{Regge}}(j)+B(j)
\end{equation}
where $S_{\text{Regge}}$ is the Regge action for a 4-simplex. One of the two exponentials in which the cosine can be decomposed suppresses the sum \eqref{somma} through the presence of a rapidly oscillating phase. The term $B(j)$ in \eqref{10jRegge} is dominant but it is suppressed inside the sum \eqref{somma} through a non trivial mechanism \cite{Bianchi:2007vf}. Expanding $S_{\text{Regge}}$ up to second order the result \eqref{amplilargej} is obtainded by a simple gaussian integration. We refer to the literature on the graviton propagator for this technique. \\
In the range $1 \lesssim j \lesssim 4$ we have interpolated the quantity (\ref{ampli}) numerically. 
Data are fitted with the function $a/j^{\alpha}$, where
$a \approx 0.66 $, and $\alpha \approx 0.73$,
\begin{eqnarray}
\eta \, \langle W| \widehat{1/A^{\rm a}} |\Psi_j \rangle  \approx \frac{{0.66 }}{j^{0.73}}.
\label{Hscale}
\end{eqnarray}
What we learnt from the explicit calculation of (\ref{ampli}) can be summarized as follows,
\begin{eqnarray}
\Big\langle \widehat{\frac{1}{A^{\rm a}}} \Big\rangle  \approx \left\{ \begin{array}{ll} 
          \frac{1}{[j(j+1)]^{\frac{1}{2}}l_P^2} & {\rm for} \,\,\, j\gg1  \,\,\, (j \gtrsim 7),\vspace{0.1cm} \\ 
         \frac{0.66}{j^{0.73}}   & {\rm for} \,\,\, 1\leq j \lesssim 7
        \end{array} \right. 
\label{Flimits}
\end{eqnarray}
The result (\ref{Flimits}) can be used to define 
an effective Laplacian (formula \eqref{pmedgb})
\begin{eqnarray}
\Delta_j  \approx  \left\{ \begin{array}{ll} 
         \frac{A_{j_0}^{{\rm a}}}{[j (j+1)]^{1/2}} \Delta_{j_0} &{\rm for} \,\,\, 7\lesssim j\lesssim j_0, 
         \vspace{0.1cm}
         \\
                  \frac{a}{j^{\alpha}} \Delta_{j_0}  & 
         {\rm for} \,\,\, j \approx 1 \,\,\, (1\lesssim j \lesssim 4),
        \end{array} \right. 
\label{deltalimits}
\end{eqnarray}
where we introduced the infrared scale 
$j_0 \gg 1$, 
 and, 
by definition, $A_{j_0}^{{\rm a}} = [j_0(j_0+1)]^{1/2}$. 
Actually (\ref{deltalimits}) is correct also for the area spectra $A_j^{\rm b}$ and $A_j^{\rm c}$
but with different values of $\alpha$, $a$ and the proper functions $A_{j_0}^{\rm b}$ and $A_{j_0}^{\rm c}$.


We introduce here a physical input to put the momentum $k$ in our 
analysis.
If we want to observe the spacetime with a microscope of resolution $l=l_P j^{1/2}$ 
(the infrared length is $\ell_0 := l_P j_0^{1/2}$) 
we must use a probing particle of momentum $p \sim 1/l$. 
The energy scaling property of the Laplacian, i.e. the scaling function $F(p)$, can be obtained 
by replacing: $l\sim 1/p$, $l_0\sim 1/p_0$ and $l_P\sim 1/E_P$, where $p_0$ is an infrared energy cutoff and $E_P$ is the Planck energy. This gives
\begin{align}
F(p)  \approx  \left\{ \begin{array}{ll} 
         \left[\frac{p^4 (E_P^2 + p_0^2)}{p_0^4 (E_P^2 +p^2)}\right]^{\frac{1}{2}}  + 1 
       &{\rm for} \,\,\,p \lesssim \frac{E_P}{\sqrt{7}}
         \vspace{0.1cm}
         \\c^{\prime } \, p^{2 \alpha}  
     & {\rm for} \,\,\, \frac{E_P}{2} \lesssim p \lesssim E_P
        \end{array} \right. 
\label{Sklimits}
\end{align}
The term $+1$ is added to facilitate the 
calculations of the spectral dimension.
Notice that the scaling function $F(p)$ 
represents also by construction 
 the scaling of the inverse metric, 
 $\langle g^{\mu \nu} \rangle_p = F(p) \langle g^{\mu \nu} \rangle_{p_0}$.

The parameter $z$ introduced in the second section for the area operator $A^{\rm a}$ is 
\begin{align}
z  \simeq  \left\{ \begin{array}{ll}2 &{\rm for} \,\,\,  p \lesssim \frac{E_P}{\sqrt{7}}
         \vspace{0.1cm}
         \\1.73 \,\, &{\rm for} \,\,\, \frac{E_P}{2} \lesssim p \lesssim E_P
        \end{array} \right. 
\label{Sklimits}
\end{align}
The dependence from $T$ in \eqref{TPKwithQG} determines the fractal dimensionality of
spacetime via (\ref{dimform}). In the limits $T\rightarrow\infty$ and 
$T\rightarrow 0$ where we are probing very large and small distances,
respectively, we obtain the dimensionalities corresponding to the largest
and smallest length scales possible. These limits are governed by the behaviour of $F(p)$ for 
$p\rightarrow 0$ and $p\rightarrow\infty$, respectively.
\begin{figure} 
 \begin{center}
 \vspace{0.5cm}
 \hspace{-0.0cm} \includegraphics[height=3.5cm]{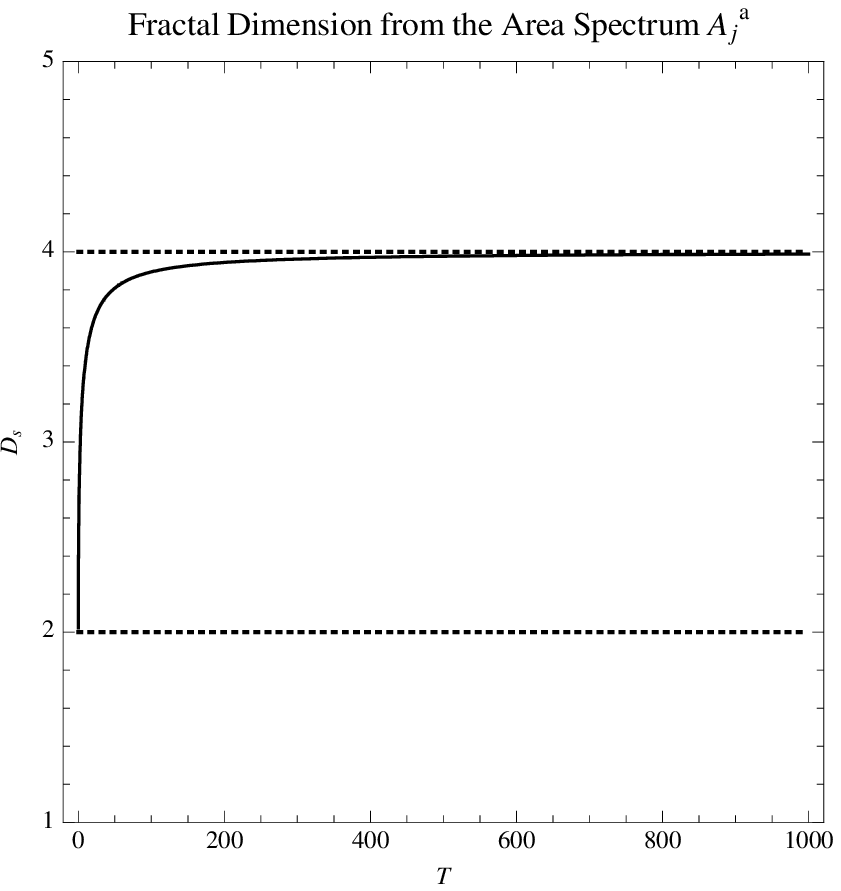} \hspace{1cm}
 \includegraphics[height=3.5cm]{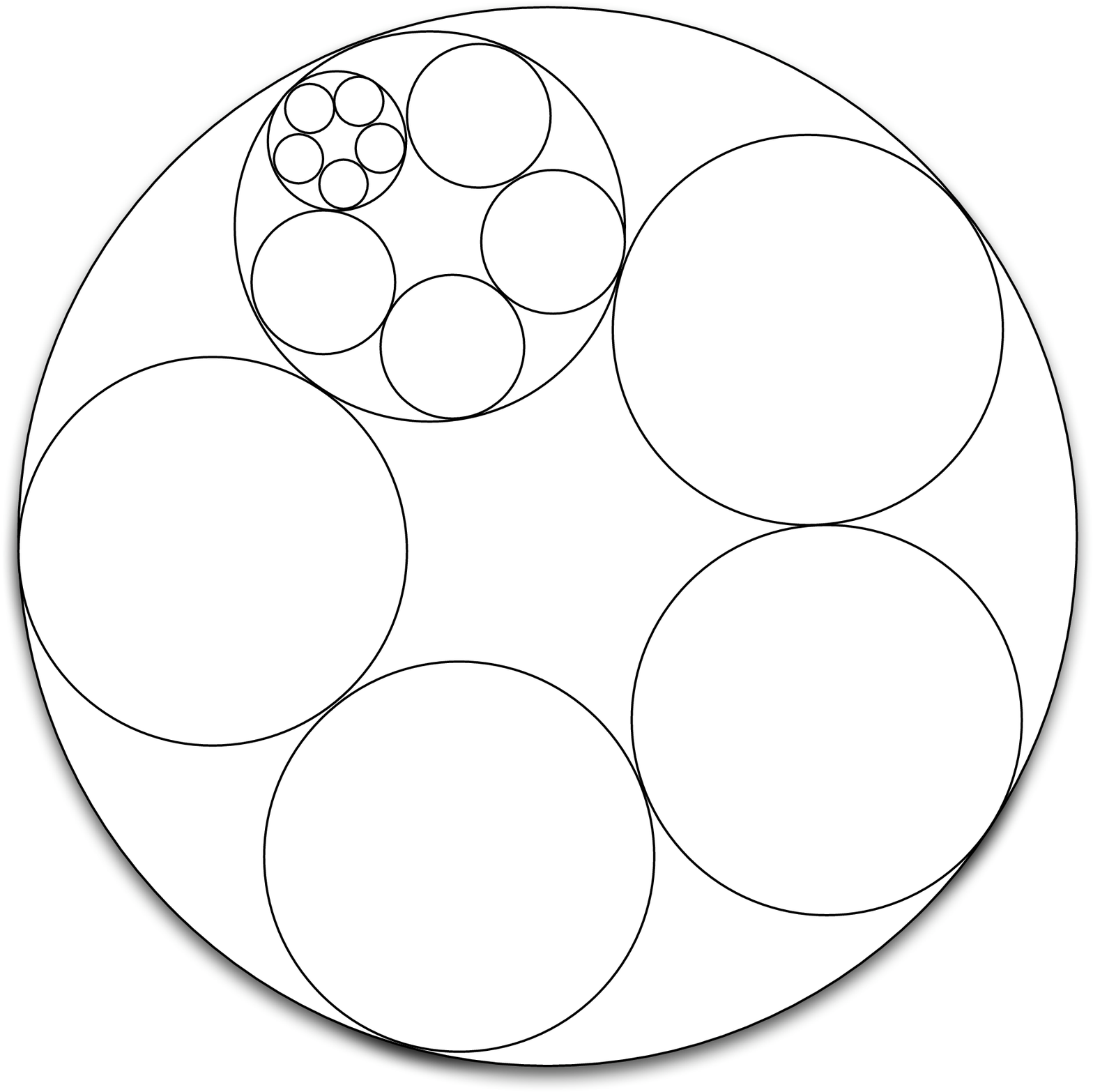}
    \end{center}
    \vspace{-0.0cm}
  \caption{\label{PlotDj6} Picture on the left:
 plot of the spectral dimension for $j\gtrsim 7$ or $k\lesssim E_P/\sqrt{7}$
 as function of the fictitious time $T$. The dimension at close to the Planck scale is $2$.
 We have plotted $T \in [0, 100]$ and used $E_p =100$, $k_0 =1$.
 Picture on the right: this is an artistic representation of the fractal properties of the spacetime.
 At any scale the spacetime appears as a ball (because we consider a compact space time but the result is independent from the spacetime topology) and the collection of all those balls (one for any scale) 
 originates a fractal quantum spacetime. 
 }
\end{figure}
The spectral dimension increases from $2$ to $4$ when the momentum of the probe field decreases from $p \approx E_P/\sqrt{7}$ to $p \approx p_0$. A plot of the fractal dimension is given in Fig.\ref{PlotDj6}. In the regime $E_P/2 \lesssim p \lesssim E_P$ the spectral dimension is ${\mathcal D}_s \approx 2.31$. We can repeat the analysis for the area spectra $A_j^{\rm b}$ and $A_j^{\rm c}$. The spectral dimension is the same for $j \gtrsim 5$ or $p\lesssim E_P/25$ but we can have differences at the Planck scale. We summarize the result in the following table,
\begin{center}
\begin{tabular}{|r|r|r|}
\hline
$  {\rm Area} \,\, {\rm Spectrum}$ \,\,\, & $z$ \,\,\,\, & \, ${\rm Fractal}  \,\, {\rm Dimension}$ \, \\
\hline
\hline
\, $A^{\rm a}_j= l_P^2 \sqrt{j(j+1)}$ \,   & \, 1.73 \, & 2.31 \\
\hline
\hline
\, $A^{\rm b}_j= l_P^2 (2j+1)$ \,  & \, 1.62 \, & 2.45 \\
\hline
\hline
\, $A^{\rm c}_j= l_P^2 j$ \, & \, 1.92 \, & 2.08  \\
\hline
\end{tabular}
\label{good}
\end{center}

{\em Correlations.} 
The 2-point correlation function of the Brownian motion over a Riemannian $d$-dimensional manifold is defined by 
\begin{eqnarray}\label{integpropag}
G(x,x^{\prime}) = \int_0^{+\infty} \text dT\, K(x, x^{\prime}; T).
\label{prop}
\end{eqnarray}
The heat kernel has an approximate expression for $x\approx x'$:
\begin{eqnarray}
 K(x, x^{\prime}; T) = \frac{1}{(4 \pi T)^{d/2}} 
 e^{-  \frac{\sigma(x, x^{\prime})}{4 T}} (1 + {\rm curv.}),
\label{kx}
\end{eqnarray}
where $\sigma(x, x^{\prime})$ is the geodesic distance and with ``curv." 
we have indicated the curvature corrections. For $x \approx x^{\prime}$ the result of the integration \eqref{integpropag} is
\begin{eqnarray}\label{correlformula}
G(x,x^{\prime}) \approx \frac{1}{ 4 \pi^{d/2}} \sigma(x, x^{\prime})^{1 - d/2} \, 
\Gamma\left(\frac{d}{2} -1\right).
\label{propd}
\end{eqnarray} 
We use the spectral dimension we found in the spin foam model as an effective dimension at high energies, hence in \eqref{correlformula} we make the substitution $d\rightarrow\mathcal D_s$. At low energy $p \approx p_0 \ll E_P$ the spectral dimension is $\mathcal D_s = 4$ and 
$G(x,x^{\prime}) \approx 1/4 \pi^2 \sigma(x, x^{\prime})$. In the energy range 
$E_P/2 \lesssim k \lesssim E_P$ where the spectral dimension is $\mathcal D_s=2+\epsilon$ 
the propagator is: $G(x,x^{\prime}) \approx  - \log[\sigma(x, x^{\prime}) \pi \epsilon]/4 \pi$. At the Planck scale instead, defining $\mathcal D_s = 2+\beta$ ($\beta = 0.31, 0.45, 0.08$ for the three different area spectrum), 
 $G(x,x^{\prime}) \approx  \Gamma(\beta/2)/(4 \pi^{1+\beta/2} \sigma(x, x^{\prime})^{\beta/2})$.

{\em Conclusions and Discussion.}
In this paper we computed the spectral dimension (${\mathcal D}_s$) of 4d quantum spacetime in the Barrett-Crane spinfoam model. We considered the simplest decomposition of spacetime and used the general boundary formalism to characterize the scaling properties of the expectation value for the traced propagation kernel. Our main observable is a conformal metric defined by $g_{\mu \nu} = \Omega^2 \, g^{o}_{\mu \nu}$, where $g^{o}_{\mu \nu}$ is a background metric.
In the quantum theory the conformal factor is quantized as a function of the area operator. We made our analysis with three kind of area spectra, obtaining different values for the spectral dimension at Planck scale (see table) but the same result 
for $E \lesssim E_P/2$ (see Fig.\ref{PlotDj6}). 

We interpret the results in the following way. At high energy the spectral dimension
is ${\mathcal D}_s \approx 2$
because the manifold presents holes typical of an atomic structure.

{\em Acknowledgements.}
Research at Perimeter Institute is supported by the Government of Canada through Industry Canada 
and by the Province of Ontario through the Ministry of Research \& Innovation.

\bibliographystyle{revtex}
\bibliography{fractalbibliography}

\end{document}